\documentstyle[11pt,amstex]{amsart}

\newtheorem{thm}{Theorem}[section]
\newtheorem{lem}[thm]{Lemma}
\newtheorem{prop}[thm]{Proposition}
\newtheorem{cor}[thm]{Corollary}
\newtheorem{rem}[thm]{Remark}

\renewcommand{\Re}{{\Bbb R}}	  

\newcommand{\aR}{\bar R}	  
\newcommand{\anabla}{\bar \nabla} 
\newcommand{\id}{I}		  
\renewcommand{\div}{\text{\rm div}}   
\newcommand{\CC}{{\cal C}}	  
\newcommand{\YY}{{\cal Y}}	  
\newcommand{\MM}{{\cal M}}	  
\newcommand{\DD}{{\cal D}}	  
\newcommand{\PP}{{\cal P}}	  
\newcommand{\Teich}{{\cal T}}	  
\newcommand{\Super}{{\cal S}}	  
\newcommand{\hE}{\widehat{E}}	  
\newcommand{\tE}{\widetilde{E}}	  
\newcommand{\genus}{\text{\rm genus}} 
\newcommand{\taumax}{\tau_{\text{\rm  max}}}  
\newcommand{\taumin}{\tau_{\text{\rm min}}}  
\newcommand{\half}{\frac{1}{2}}
\newcommand{\tr}{\text{\rm tr}}

\newcommand{\la}{\langle}
\newcommand{\ra}{\rangle}
\newcommand{\tens}{\otimes}	  
\newcommand{\Lie}{{\cal L}}	  
\newcommand{\Area}{\text{\rm Area}}   

\newcommand{\ame}{\bar g}	  
\newcommand{\Lapse}{N}		  
\newcommand{\Shift}{X}		  
\newcommand{\norm}{\text{\rm T}}  
\newcommand{\TT}{\text{\scriptsize TT}}
\newcommand{\CosmC}{\Lambda}	  

\newcommand{\commentout}[1]{}	 

\title{On the global evolution problem in 2+1 gravity}
\author[L. Andersson]{Lars Andersson$^1$}
\address{Department of Mathematics\\
Royal Institute of Technology\\
100 44 Stock\-holm, Sweden}
\thanks{$^1$Supported in part by the Swedish Natural
Sciences Research Council (SNSRC),  contract no.  F-FU 4873-307.} 
\email{larsa\char'100math.kth.se}

\author[V. Moncrief]{Vincent Moncrief$^2$}
\address{Department of Mathematics and Department of Physics\\
Yale University\\
P.O. Box 208120\\
New Haven, CT 06520, USA}
\email{MONCRIEF\char'100yalph2.physics.yale.edu}
\thanks{$^2$Supported in part by the NSF, contract no.
PHY-9503133}

\author[A. J. Tromba]{Anthony J. Tromba}
\address{Department of Mathematics\\
University of California Santa Cruz\\
Santa Cruz, CA 95064-1099, USA}
\email{tromba\char'100cats.ucsc.edu}

\date{Oct. 7, 1996\\
To appear in an issue of {\em Journal of Geometry and Physics} in honor of 
Andr\'e Lichnerowicz.}
\subjclass{Primary 83C50, Secondary 53C50, 30F99.}

\begin{document}

\begin{abstract}
Existence of global CMC foliations of constant curvature 3-dimensional
maximal globally hyperbolic Lorentzian manifolds, containing a constant mean
curvature hypersurface with
$\genus(\Sigma) > 1$ is proved. Constant curvature 3-dimensional
Lorentzian manifolds can be viewed as solutions to the 2+1 vacuum Einstein 
equations with a
cosmological constant. The proof is based on the reduction of
the corresponding Hamiltonian system in constant mean curvature gauge to a
time dependent Hamiltonian system on the cotangent bundle of Teichm\"uller
space. Estimates of the Dirichlet energy of the induced metric play an
essential role in the proof.
\end{abstract}

\maketitle

\section{Introduction}
Andr\'e Lichnerowicz \cite{lichnerowicz} used the conformal transformation 
properties of the scalar curvature to write the constraint equations
on Cauchy data for the Einstein equations as a semilinear
elliptic system. This important insight together with the fact that in 
the case of constant mean curvature (CMC) data the Hamiltonian and the 
momentum constraint equations decouple, leads to the conformal method for
solving the constraint equations and to the conformal method of 
reduction of the Einstein equations in CMC gauge. 

The Einstein equations in 3+1 dimensions on a spacetime of topology 
$\Sigma \times I$ for some interval $I$ reduce, using the conformal method 
(ignoring technical difficulties) to a time dependent Hamiltonian system on 
$T^*\Super(\Sigma)$, 
where $\Super(\Sigma)$ is the ``conformal  superspace'' of $\Sigma$, i.e.
the space of Riemannian metrics on $\Sigma$ modulo conformal rescalings and 
diffeomorphisms. 
In the 2+1 dimensional case with $\Sigma$ a compact oriented surface, the above
ideas can be carried out in full detail \cite{moncrief:2+1:reduction}. 

In this paper we address the evolution problem for 2+1 dimensional vacuum
gravity with a cosmological constant in CMC gauge. We prove that all 2+1
dimensional vacuum spacetimes $(\Sigma \times I, \ame)$  satisfying the vacuum Einstein equations with
cosmological constant,
$$
\aR_{ab} = \CosmC \ame_{ab},  
$$
which contain a CMC hypersurface are globally foliated by CMC hypersurfaces.
In this case equation
(\ref{eq:lambda}) below corresponds to the Lichnerowicz equation. 

Vacuum 2+1 dimensional spacetimes with a cosmological constant are 3 dimensional
Lorentzian spaceforms. The problem of classifying the
3 dimensional maximal globally hyperbolic Lorentzian spaces of constant
curvature, and of topology $\Sigma \times \Re$ was solved by Mess in
\cite{mess:const:curv} for $\CosmC \leq 0$. The question of existence of a 
global CMC foliation was left open in this work.

\subsection{Overview:}
\label{sec:overview} This paper is organized as follows.
In section \ref{sec:teichmuller} we review some facts from Teichm\"uller
theory and define the Dirichlet energy $\hE$. 
Here $\hE$ is the form of the Dirichlet energy with fixed target space 
introduced by Tromba \cite{tromba:Teich}. The fact that this is a proper 
function on Teichm\"uller space is a key ingredient in our argument.
In section \ref{sec:vac2+1} we
introduce the 2+1 Einstein evolution equations in CMC gauge and derive the 
Calabi-Simons identity (\ref{eq:CSbasic}). Since we are dealing with
constant curvature spacetimes the Riccati equation can be explicitly
solved. This is done in subsection \ref{sec:Gauss}, where for $\CosmC \geq
0$ we use this to study the asymptotics of the mean curvature of 
the Gauss foliation. 
In section \ref{sec:global} we state and prove the main results, Theorem
\ref{thm:global} (global existence) and Corollary \ref{cor:glob} (global
CMC foliations). 

The conformal constraint equations are considered in 
subsection \ref{sec:confconstr}. In subsection \ref{sec:est-evol}, some
estimates for the area and the Dirichlet energy are derived.
An application of the Calabi--Simons identities, Lemma \ref{lem:CS},
for the second fundamental form together with the maximum
principle yields pointwise bounds on the second fundamental form $K$ in terms
of the mean curvature $\tau$. 
This bound together with the Einstein evolution equations allows one to derive 
a differential inequality for the time development of the Dirichlet energy 
$\hE(g(\tau))$ of the metric $g(\tau)$ which shows that $\hE(g(\tau))$ is
finite for all $\tau$ allowed by the constraint equations.
Finally
in subsection \ref{sec:proof-global} the proof of the main theorem is
carried out, using the bound on the Dirichlet energy derived in subsection
\ref{sec:est-evol}.

\section{Teichm\"uller Space}
\label{sec:teichmuller}
In this section we present some background material on Teichm\"uller space. The
main reference is the book by Tromba \cite{tromba:Teich}.
Fix a compact oriented 2-manifold $\Sigma$ of genus $> 1$.
Let $\MM$ denote the space of $C^{\infty}$ metrics on $\Sigma$,
let $\MM_{-1} \subset \MM$ denote the space of metrics of constant
scalar curvature $-1$,
let $\DD_0$ denote the group of $C^{\infty}$ diffeomorphisms of $\Sigma$
isotopic to the identity and let $\PP$ be the group of positive $C^{\infty}$
functions on $\Sigma$.

The scalar curvature function obeys the following
transformation rule in two dimensions,
\begin{equation}\label{eq:confR}
R[e^{2\lambda} g ] = e^{-2\lambda}(-2 \Delta_g \lambda	+ R[g]) .
\end{equation}
The equation
$$
\Delta_g \lambda = \half ( R[g] + e^{2\lambda} )
$$
has a unique solution in case $\genus(\Sigma) > 1$ and therefore
we may construct a metric $h = e^{2\lambda} g \in \MM_{-1}$ conformal to  $g$
and with constant scalar curvature $-1$.
$\PP$ acts on $\MM$ by conformal rescaling and
$\MM_{-1}$ is a global slice for the action of $\PP$. Thus
$$
\MM_{-1} = \MM / \PP .
$$
Any $C^{\infty}$ symmetric 2-tensor $k$ can be decomposed uniquely as
\begin{equation}\label{eq:york-decomp}
k = k_{\TT} + f g + L_g(Y) ,
\end{equation}
where $k_{\TT}$ is transverse traceless, i.e. $\tr_g k_{\TT} = 0, \div_g
k_{\TT} = 0$, $f \in C^{\infty}(\Sigma)$ and $L_g(Y)$ is the conformal Killing
form for the vector field $Y$, i.e. the trace free part of $\Lie_Y g$,
$$
L_g(Y) = \Lie_Y g - \frac{1}{2} \tr_g(\Lie_Y g) g .
$$
The decomposition (\ref{eq:york-decomp}) is $L^2$ orthogonal. In 2 dimensions
$\div$ acting on
traceless symmetric 2-tensors is elliptic and by Riemann--Roch the dimension
of the kernel has dimension $6\genus(\Sigma)-6$ for $\genus(\Sigma) >1$. 
Further in 2 dimensions, transverse traceless is a conformally invariant 
property.
We will use the notation $S^2_{\TT}(T^*\Sigma,h)$ for the space of
$\TT$-tensors w.r.t. $h$.

Let $k \in T_h \MM_{-1}$. Then $k$ is of the form $k_{\TT} + \Lie_X h$ for
some $X$. The $\TT$ tensors provide a local slice for
the action of $D_0$ on $\MM_{-1}$. The action of $\DD_0$ is proper
so Teichm\"uller space $\Teich(\Sigma)$ defined by
$$
\Teich(\Sigma) = \MM_{-1} / \DD_0 .
$$
is a manifold of dimension $6\genus(\Sigma)-6$.

$\MM$ and $\MM_{-1}$ are Riemannian manifolds w.r.t. the $L^2$-metric defined
for $h,k \in T_g\MM$ by
$$
\la\la h,k \ra\ra_g = \int_{\Sigma} \la h,k\ra_g \sqrt{g} d^2x
$$
and $\half \la\la \cdot, \cdot \ra\ra$ restricted to $\MM_{-1}$ induces a
Riemannian structure $\la \cdot , \cdot \ra_{WP}$ on
$\Teich(\Sigma)$, the Weil-Peterson metric.

We need some basic facts about harmonic mappings. Let $G$ be a fixed
element in $\MM_{-1}$ and for $g \in \MM_{-1}$, $S : \Sigma \to \Sigma$,  let
$$
e(S,g) = \half g^{ij}
\frac{\partial S^{\alpha}}{\partial x^i}
\frac{\partial S^{\beta}}{\partial x^j} G_{\alpha\beta}
= \half \la \nabla S^{\alpha} , \nabla S^{\beta} \ra_g G_{\alpha \beta}
$$
and let $E(S,g)$ be given by
\begin{equation}\label{eq:E(S)}
E(S,g) = \int_\Sigma e(S,g) \sqrt{g} d^2 x .
\end{equation}
There is a unique map homotopic to the identity which minimizes
$E(S,g)$. This map, which we denote by $S(g)$, is a harmonic map 
$(\Sigma,g) \to (\Sigma,G)$ and it can be proved that $S(g) \in \DD_0$.

Let $\sigma: \MM_{-1} \to \MM_{-1}$ be defined by
\begin{equation}\label{eq:sigmadef}
\sigma(g) = S(g)_* g .
\end{equation}
Since $S(g) \in \DD_0$ the pushforward is well defined. By the
uniqueness of $S(g)$ we have $\sigma(f^* g) = \sigma(g)$ for any
$f \in \DD_0$. Therefore $\sigma$ induces a map
\begin{equation}\label{eq:sigmateich}
\sigma: \Teich(\Sigma) \to \MM_{-1}
\end{equation}
which is a global slice for the action of $\DD_0$,
see \cite[\S 3.4]{tromba:Teich}. By a slight abuse of notation we will say
that $h \in \sigma$ if $h \in \sigma(\Teich(\Sigma)) \subset \MM_{-1}$.
Given $h \in \MM_{-1}$ or $g \in \MM$, we denote the corresponding classes in
$\Teich(\Sigma)$ by $[h]$ and $[g]$.

We summarize the relevant facts in
\begin{prop}[\cite{tromba:Teich}] \label{prop:teichbasic}
\ \
\begin{enumerate}
\item $\MM_{-1} = \MM / \PP $ and $\MM_{-1}$ is a global slice for the
action of $\PP$.
\item \label{6g-6}
$\Teich(\Sigma) = \MM_{-1} / \DD_0$ and there is a global slice for the action
of $\DD_0$ given by the map $\Teich(\Sigma) \to \MM_{-1}$ induced by the
diffeomorphism invariant map $\MM_{-1} \to \MM_{-1}$ given by
$\sigma(g) = S(g)_* g$, with $S(g) \in \DD_0$ the harmonic  minimizer of
(\ref{eq:E(S)}).
$\Teich(\Sigma)$ is a $C^{\infty}$ manifold of dimension
$\dim \Teich(\Sigma) = 6 \genus(\Sigma) -6$ diffeomorphic to
$\Re^{6 \genus(\Sigma) -6}$
\item The weak Riemannian metric on $\MM_{-1}$ induces a
metric $\la \cdot , \cdot \ra_{WP}$ on $\Teich(\Sigma)$, the Weil-Peterson 
metric.  $\Teich(\Sigma)$ is geodesically convex w.r.t. the Weil-Peterson 
metric but not complete.
\end{enumerate}
~\qed
\end{prop}

We define the Dirichlet energy of $g \in \MM_{-1}$ to be
\begin{equation}\label{eq:E(g)}
E(g) = E(S(g),g)
\end{equation}
$E(g)$ turns out to be conformally
invariant and diffeomorphism invariant. By the conformal invariance, $E(g)$
extends to a function on $\MM$ which we denote by
$\hE$.
Further, by the diffeomorphism invariance of $E$ it defines a
function $\tE$ on $\Teich(\Sigma)$ which we again call the Dirichlet energy.

\begin{prop}[\cite{tromba:Teich}]\label{prop:tromba}
\ \
\begin{enumerate}
\item
The Dirichlet energy $\tE$ is a proper function on $\Teich(\Sigma)$.
\item
Let $g \in \MM$ and let $S(g) \in \DD_0$ be the harmonic map as in
(\ref{eq:E(g)}).
For $h \in T_g(\MM)$, let $\tilde h^{\sharp}$ denote the
(1,1)--version of the trace free part of $h$. Then
\begin{equation}\label{eq:DhE}
D \hE(g) h =  -\half \int_\Sigma \la \tilde h^{\sharp} \nabla S^{\ell} ,
\nabla S^{\ell}	 \ra_g \sqrt{g} d^2 x .
\end{equation}
\end{enumerate}
~\qed
\end{prop}

\section{Vacuum 2+1 gravity}\label{sec:vac2+1}
Assume that $\Sigma$ is a compact and orientable 2-dimensional manifold
and let
$(\Sigma \times I, \ame)$ be a maximal globally hyperbolic 2+1 spacetime, which
solves the Einstein vacuum equations with cosmological constant $\CosmC$.
We will denote the covariant derivative and curvature tensor defined w.r.t.
$\ame$ by $\anabla, \aR$, respectively.
The field equations are
\begin{equation}\label{eq:2+1field}
\aR_{ab} = \CosmC \ame_{ab}  .
\end{equation}
Note that equation (\ref{eq:2+1field}) due to the 3-dimensionality of
$\Sigma \times I$
is equivalent to
$$
\aR_{abcd} = \frac{\CosmC}{2} (\ame_{ac} \ame_{bd} - \ame_{ad} \ame_{bc} ) ,
$$
i.e. $(\Sigma \times I, \ame)$ is a Lorentzian spaceform of sectional curvature
$\CosmC/2$.

We assume that the constant time hypersurfaces $\Sigma_t = \Sigma \times
\{t\}$ are spacelike with normal $\norm$
and denote the induced metric by $g$. Define the lapse function $\Lapse$ and
shift vectorfield $\Shift$ by
$$
\Lapse = - \la	\partial_t , \norm \ra , \qquad
\Shift = \Shift^a \partial_{x^a} = \partial_t - \Lapse \cdot \norm .
$$
Then we can write $\ame$ in the form
\begin{equation}\label{2+1metr}
\ame = - \Lapse^2 dt \tens dt 
+ g_{ab} (dx^a + \Shift^a dt )\tens (dx^b + \Shift^b dt) .
\end{equation}
Let $K$ denote the
second fundamental form, i.e.
$$
K_{ab} = - \la \anabla_{a} \norm, e_b\ra = \la \norm , \anabla_{a} e_b \ra
$$
and let
\begin{equation}\label{eq:piK}
\pi^{ab} =  K_c^{\ c} g^{ab} - K^{ab} ,
\end{equation}
then ${\pi'}^{ab} = \sqrt{g} \pi^{ab}$ is the canonically conjugate variable to
$g$ in the Hamiltonian formulation of Einsteins equations.
Note that in \cite{moncrief:2+1:reduction} the notation $\pi^{ab}$ rather
than ${\pi '}^{ab}$ was used.

In the 2+1  case with cosmological constant $\CosmC$ we have from the Gauss
and Gauss-Codazzi equations
\begin{eqnarray}
(\tr_g \pi)^2 - |\pi|_g^{2} + R &=& \CosmC ,
\label{eq:CosmHamconstr1}\\
\nabla_a \pi^{ab} &=& 0 . \label{eq:CosmHamconstr2}
\end{eqnarray}
Using the second variation equations and the definition of $\Lapse, \Shift$
one arrives after a bit of calculation at the equations of motion which
in terms of  $(g,\pi)$ are
\begin{eqnarray}
\partial_t g_{ab}  &=& 2\Lapse(\pi_{ab} -(\tr_g\pi)  g_{ab})
+ (\Lie_X g)_{ab} ,  \label{eq:dtgpi} \\
\partial_t \pi_{ab} &=& \nabla_a \nabla_b N - \Delta N g_{ab}
- N \frac{\CosmC}{2}  g_{ab}  \label{eq:dtpi}\\
&& \hskip 0.2in + N[ |\pi|_g^2 g_{ab} - (\tr_g \pi)^2 g_{ab} + \pi_{ac}
\pi^c_{\ b} ] + (\Lie_X \pi)_{ab} . \nonumber
\end{eqnarray}

Let $\tau = \tr_g \pi$.
We will in the following consider only the case of constant mean curvature
(CMC) data, with $d\tau = 0$.
The lapse function corresponding to a CMC slicing satisfies
\begin{equation}\label{eq:CMCcond}
 - \Delta \Lapse + ( |\pi|_g^2 - \CosmC) \Lapse = 1 .
\end{equation}
By (\ref{eq:york-decomp}) and the momentum constraint equation
(\ref{eq:CosmHamconstr2}), $\pi$ satisfies
\begin{equation}\label{pipiTT}
\pi = \frac{\tau}{2} g + \pi_{\TT} , \qquad |\pi|_g^2 = \frac{\tau^2}{2} +
|\pi_{\TT}|_g^2 ,
\end{equation}
in the CMC case. 

In case $\Sigma \cong S^2$, then $\Teich(\Sigma)$ is zero dimensional and
there are no nontrivial $\TT$-tensors.
It follows from this and the constraint equations
(\ref{eq:CosmHamconstr1}--\ref{eq:CosmHamconstr2}) that for $\Sigma \cong
S^2$, the 2+1 vacuum Einstein
equations have solutions only in case
$\CosmC > 0, \tau^2 < 2 \CosmC$ and the solution in this case is given by
\begin{equation}\label{eq:trivsol}
g_{ab}(\tau) = g_{ab}(\tau_0) \frac{\tau_0^2 - 2\CosmC}{\tau^2 - 2\CosmC} .
\end{equation}
This is the 2+1 dimensional deSitter universe.

If $\Sigma \cong T^2$, any $\TT$-tensor is
covariant constant and the space of $\TT$-tensors is 2-dimensional. The
equations of motion can be explicitly solved.

For $\genus(\Sigma) > 1$ we do not have an explicit solution of the equations of
motion except in the trivial case $\pi_{\TT} = 0$. In this case, the
solutions of the field equations can be described as follows. 
If $\genus(\Sigma) > 1$ and $\CosmC \geq 0$, the constraint equations 
imply  $\tau^2 > 2 \CosmC$ and we may therefore assume
$\tau_0 > \sqrt{2\CosmC}$. The evolution of trivial data is given by
(\ref{eq:trivsol}).
We see that in the case when $\CosmC \geq 0$, the trivial solutions undergo
an infinite expansion as $\tau \searrow \sqrt{2\CosmC}$, collapse to a 
singularity as $\tau \to \infty$, while when $\CosmC < 0$,
$\tau$ runs from $-\infty$ to $+\infty$ and  we have a ``big bang'' and a
``big crunch''.
In case $\CosmC =0$, the trivial solutions correspond to
quotients of the interior of a light-cone in the 2+1 dimensional Minkowski
space, while in case $\CosmC < 0$, the trivial solutions correspond to
quotients of a maximal globally hyperbolic subset of the 2+1 dimensional
anti-deSitter space.

\begin{lem}\label{lem:CS}
The following identities hold for $K \in S^2_{\TT}(T^*\Sigma,g)$.
\begin{eqnarray}
\label{eq:codazzi}
\nabla_c K_{ab} - \nabla_b K_{ac} &=& 0 , \\
\label{eq:CSbasicnew}
\nabla^c \nabla_c K_{ab} &=& R K_{ab} , \\
\label{eq:CSbasic}
\half \Delta |K|_g^2
&=& | \nabla K|_g^2 + R |K|_g^2 .
\end{eqnarray}
\end{lem}
\begin{pf}
To prove the identity (\ref{eq:codazzi}) note that
if $K \in S^2_{\TT}(T^*\Sigma,g)$ then in an ON frame
\begin{eqnarray*}
K_{11;1} + K_{12;2} &=& 0 , \\
K_{21;1} + K_{22;2} &=& 0 , \\
K_{11;1} + K_{22;1} &=& 0 , \\
K_{11;2} + K_{22;2} &=& 0 ,
\end{eqnarray*}
which after an elementary manipulation gives (\ref{eq:codazzi}).
Let $K \in S^2_{\TT}(T^*\Sigma,g)$. A computation in an ON frame gives
\begin{eqnarray*}
K_{ij;kk} &=& K_{ik;jk} \\
K_{ij;kk} &=& K_{ik;jk} \\
&=& K_{ik;kj} + R_{fijk} K_{fk} + R_{fkjk} K_{if}
\end{eqnarray*}
By (\ref{eq:codazzi}) and $K \in S^2_{\TT}(T^*\Sigma,g)$,
$K_{ik;kj} = K_{kk;ij} = 0$ and using
the fact that in two dimensions, the Riemann
tensor is of the form
$$
R_{abcd} = \half R ( g_{ac}g_{bd}  - g_{ad}g_{bc} ) 
$$
gives (\ref{eq:CSbasicnew}). Finally, (\ref{eq:CSbasic}) is proved from
(\ref{eq:CSbasicnew}) by a straightforward computation.
\end{pf}
\begin{rem}
Note that the identities (\ref{eq:codazzi})--(\ref{eq:CSbasicnew}) are special
to two dimensions. Equation (\ref{eq:CSbasic}) is one of the Calabi-Simons
identities, analoques of this hold for Codazzi tensors in higher dimensions,
and for higher covariant derivatives of $K$,
see \cite{choi:treibergs:gauss} and references therein.
The above identity may also be proved by differentiating the 2 dimensional
identity
$R_{ab} - \half R g_{ab} = 0$ in the direction of a $\TT$ tensor.
~\qed
\end{rem}

The Calabi-Simons identity enables us to apply the maximum principle to
get pointwise estimates for $\pi_{\TT}$ and $\Lapse$.
\begin{lem}\label{lem:Lapsepiest} Assume $\genus(\Sigma) > 1$.
Let $(g_{ab},\pi^{ab})$ be CMC data with mean curvature $\tau$,
for 2+1 vacuum GR with cosmological constant $\CosmC$ and
let $\Lapse$ be the lapse function. Then $\tau^2/2 > \CosmC$ and
\begin{eqnarray}
|\pi_{\TT}|_g^2 &\leq& \frac{\tau^2 - 2\CosmC}{2}  ,  \label{eq:tKest}\\
\frac{1}{\tau^2 - 2\CosmC} &\leq& \Lapse \leq \frac{2}{\tau^2 - 2 \CosmC} .
\label{eq:Lapseest}
\end{eqnarray}
\end{lem}
\begin{pf}
Let $\pi_{\TT}$ be the $\TT$ part of $\pi$ and apply Lemma \ref{lem:CS}.
The maximum principle and (\ref{eq:CSbasic})
implies that at a maximum of $|\pi_{\TT}|_g^2$,
$$
0 \geq R .
$$
Using (\ref{pipiTT}) and the Hamiltonian constraint
(\ref{eq:CosmHamconstr1})
proves (\ref{eq:tKest}). It follows that $\tau^2 \geq 2\CosmC$, but equality
implies $R = 0$ which is ruled out by $\genus(\Sigma) >1$.
Similarly, applying the maximum principle using (\ref{eq:CMCcond})
proves (\ref{eq:Lapseest}).
\end{pf}
Note that in the case when $\CosmC \geq 0$, the range of mean curvatures $\tau$
is limited by $2 \CosmC < \tau^2 < \infty$.

\subsection{Gauss coordinates} \label{sec:Gauss}
Before studying the evolution in the CMC gauge, it is instructive to consider
the geometry of the spacetime in Gauss coordinates. Due to the fact that
$(\Sigma \times I, \ame)$ has constant curvature, we can integrate the Riccati
equation.

Let $f: \Sigma\times \Re^+ \to \Sigma \times I$
be the normal exponential map, defined by
$f_s(x) = \exp_x(s T)$ and let $F(s)$ be defined by
$$
F(s) X = f_{s*} X ,
$$
where $f_{s*}$ denotes push forward under $f_s$. Then $F$ satisfies
the Jacobi equations
\begin{equation}\label{eq:jacobi}
\ddot F + \aR_T F = 0 ,
\end{equation}
where $\aR_T$ is defined by $\aR_T X = \aR(X,T)T$, i.e. in the case of an
2+1-dimensional spacetime of constant curvature $\CosmC/2$, we have
$$
\aR_T = -\frac{\CosmC}{2} \id .
$$

Let $S$ be the Weingarten map given by $SX = - \nabla_X T$ where $T$ is the
timelike normal to $f_s(\Sigma)$. Then $S = - \dot F F^{-1}$ and
$S$ satisfies the
Riccati equation
\begin{equation}\label{eq:riccati}
\dot S = S^2 + \aR_T .
\end{equation}
In the constant curvature case, equation (\ref{eq:jacobi}) has with
$F(0) = I, \dot F(0) = -S_0$,
the solution
\begin{eqnarray}
F(s) &=& \id - s S_0,\quad \text{for $\CosmC = 0$} ,  \label{eq:jacobisol1}\\
F(s) &=& \id \cosh(\sqrt{\frac{\CosmC}{2}} s)
-  \sqrt{\frac{2}{\CosmC}} S_0 \sinh(\sqrt{\frac{\CosmC}{2}} s)
, \quad \text{for $\CosmC > 0$} , \label{eq:jacobisol2} \\
F(s) &=& \id \cos(\sqrt{\frac{|\CosmC|}{2}} s)
- \sqrt{\frac{2}{|\CosmC|}} S_0 \sin( \sqrt{\frac{|\CosmC|}{2}} s)
, \quad \text{for $\CosmC < 0$}	 \label{eq:jacobisol3}.
\end{eqnarray}

Let $\genus(\Sigma) > 1$ and assume that $(g_0, S_0)$ are the metric and
Weingarten map of a hypersurface $\Sigma_{\tau_0}$ with constant mean curvature
$\tau_0$ in a
2+1 vacuum spacetime with cosmological constant $\CosmC$.
Recall that $S$ is just the (1,1)-form of the second fundamental form $K$.
It follows from (\ref{eq:tKest}) that in case $\CosmC \geq 0$, 
after a choice of time orientation, 
$$
S_0 \leq \frac{-\tau_0 + \sqrt{ \tau_0^2 -2\CosmC} }{2} I  < 0 .
$$
In case $\CosmC \geq 0$ we find from
(\ref{eq:jacobisol1}--\ref{eq:jacobisol2}) that the solution to the Jacobi
equation (\ref{eq:jacobi}) exists and is nondegenerate for all $s> 0$ and hence
$\Sigma_{\tau_0}$ has no focal points in the expanding direction.
A computation shows that 
$$
\lim_{s \to \infty} \tr S(s) = - \sqrt{2\Lambda}.
$$

Finally, we note that in case $\CosmC \geq 0$ we have causal geodesic 
completeness in the expanding direction. To see this, use that 
by global hyperbolicity and 
the fact that the time variable $s$ for the Gauss 
foliation is proper time, any point in $\Sigma \times I$ which is in the 
expanding direction w.r.t. $\Sigma_{\tau_0}$ is on one of the leaves in 
the Gauss foliation and hence cannot be an endpoint for an 
inextendible causal geodesic.
\section{Global existence}
\label{sec:global}
We now state the main theorem
\begin{thm}\label{thm:global}
Fix a compact oriented 2-manifold $\Sigma$ of genus $> 1$.
Let $(\Sigma \times I,\ame)$ be a globally hyperbolic spacetime solving the
vacuum Einstein equations {\rm (\ref{eq:2+1field})} with cosmological constant
$\CosmC$, which is the maximal globally hyperbolic development of CMC data
$(g,\pi)$ on $\Sigma$ with mean curvature $\tau_0$. If $\CosmC \geq 0$,
assume that
$\tau_0 > \sqrt{2\CosmC}$.  Then the following is true:
\begin{enumerate}
\item \label{point:glob1}
The Einstein evolution equations with CMC time gauge and spatial gauge
given by the slice $\sigma$ of Point \ref{6g-6} of Proposition
\ref{prop:teichbasic} has solution for all $\tau$ allowed by the constraint 
equations, i.e.  for
\begin{eqnarray*}
\sqrt{2\CosmC} &<& \tau < \infty, \qquad \CosmC \geq 0, \\
-\infty &<& \tau < \infty, \qquad \CosmC < 0 . 
\end{eqnarray*}
\item \label{point:glob2}
The area $\Area(\Sigma, g(\tau))$ of $\Sigma$ w.r.t. the induced
metric $g(\tau)$ at mean curvature time $\tau$ satisfies
$\Area(\Sigma, g(\tau)) \to 0$ as $\tau \to \pm \infty$ and in the case
$\CosmC \geq 0$,
$\Area(\Sigma, g(\tau)) \to \infty$ as $\tau \searrow \sqrt{2\CosmC}$.
\item\label{point:glob3}
In case $\CosmC \geq 0$, for $\sqrt{2\CosmC} < \tau < \tau_0$, 
$$
\hE(g(\tau)) \leq \hE(g(\tau_0))
\left (
\frac{\tau_0 + \sqrt{ \tau_0^2 - 2 \CosmC}}{\tau + \sqrt{\tau^2 - 2\CosmC}}
\right )^{\sqrt{2}} \!\!\!\!\!\!.
$$
In particular, for $\CosmC > 0$, the Dirichlet energy $\hE(g(\tau))$ is
bounded for the evolution in the expanding direction
$\tau \searrow \sqrt{2\CosmC}$
and the class of $ g(\tau) $ stays in a compact subset of
$\Teich(\Sigma)$.
\end{enumerate}
\end{thm}

Given the global existence for the evolution in CMC time we are now able to
prove that the spacetime is globally foliated by CMC hypersurfaces.
\begin{cor}\label{cor:glob}
Let $(\Sigma \times I, \ame)$ be as in Theorem \ref{thm:global}. Then 
$(\Sigma \times I, \ame)$ is globally foliated by  CMC hypersurfaces.
~\qed
\end{cor}
\begin{pf}
Let $\Sigma_{\tau}$ denote a CMC hypersurface with mean curvature $\tau$ as
constructed in Theorem \ref{thm:global}.
First we consider the case when $\tau \to \pm \infty$, i.e. the 
collapsing direction. By a choice of time orientation it is sufficient to
consider the case $\tau \nearrow \infty$. If we can show that the focal 
distance along future directed normal geodesics to the CMC hypersurface 
$\Sigma_{\tau}$ tends to $0$ as $\tau \nearrow \infty$ then it follows 
by global hyperbolicity that 
the CMC foliation constructed in Theorem \ref{thm:global} exhausts the 
spacetime in the collapsing direction. 
Let $F$ be defined as in subsection \ref{sec:Gauss} by solving the Jacobi
equation w.r.t. future directed normal geodesics to $\Sigma_{\tau}$ for 
some large $\tau$. Focal points correspond precisely to zero 
eigenvalues of $F$. Using the explicit form of $F$ given by
(\ref{eq:jacobisol1}--\ref{eq:jacobisol3}) and the fact that at least one of
the eigenvalues of the Weingarten map $S$ of $\Sigma_{\tau}$ is larger 
than $\tau_0/2$, one shows easily that the focal distance tends to zero as 
$\tau_0 \nearrow \infty$. 

Next we consider the case $\CosmC \geq 0$ and $\tau \searrow \sqrt{2\CosmC}$. 
Fix some $\tau_0$ satisfying the conditions of Point \ref{point:glob1} of 
Theorem \ref{thm:global}.
Using (\ref{eq:Lapseest}) and the definition of the lapse function 
we see that the Lorentz distance from $\Sigma_{\tau}$ to any point in the past
of $\Sigma_{\tau_0}$ will decrease to zero before $\tau$ reaches
$\sqrt{2\CosmC}$. Therefore the CMC foliation for 
$\sqrt{2\CosmC} < \tau < \tau_0$ exhausts the past of $\Sigma_{\tau_0}$. 
\end{pf}

\begin{rem} In the case $\CosmC \leq 0$, the case when $\tau \to \pm \infty$
in the proof of Corollary \ref{cor:glob} is covered by a standard argument 
which shows that in the case of a crushing singularity, the CMC hypersurfaces
foliate a neighborhood of the singular boundary
if the strong energy condition holds. The basic 
comparison argument used for the proof of this can easily be adapted 
to cover the present situation. 
\end{rem}

\subsection{The conformal constraint equations}\label{sec:confconstr}
Here we review the conformal procedure for solving the constraint equations, 
which is an essential step in the reduction of vacuum 2+1 gravity to a 
Hamiltonian system on $T^* \Teich(\Sigma)$, the cotangent bundle of 
Teichm\"uller space, see \cite{moncrief:2+1:reduction}.

The natural phase space for relativity is $T^*\MM$, the cotangent bundle of
the space of metrics. As discussed above the fiber in $T^* \MM$ consists of
contravariant symmetric 2-tensor densities. For simplicity of notation we
will work with tensors here.

Let $\CC_{\tau,\sigma}$ denote the space of solutions of the constraint
equations (\ref{eq:CosmHamconstr1})--(\ref{eq:CosmHamconstr2}) over the slice
$\sigma$ given by (\ref{eq:sigmateich}),
such that $(g,\pi) \in \CC_{\tau,\sigma}$ if and only if $(g,\pi)$ solves the
constraint equations, $\tr_g \pi = \tau$ and $g = e^{2\lambda} h$ for
$h \in \sigma$. Then $\CC_{\tau,\sigma}$ inherits a symplectic structure from
the $L^2$ symplectic structure on $T^* \MM$.

We will construct a map
$$
\YY_{\tau,\sigma} : T^*\Teich(\Sigma) \to \CC_{\tau,\sigma} .
$$
Let $h \in \MM_{-1}$ and let $\omega \in S^2_{\TT}(T^*\Sigma,h)$.
Then, letting
\begin{equation}\label{eq:confg}
g = e^{2\lambda} h
\end{equation}
we also have $\omega \in S^2_{\TT}(T^*\Sigma,g)$ by the two
dimensionality of $\Sigma$. From this we get that
$\pi_{\TT}^{ab} = e^{-4\lambda} p_{TT}^{ab}$ is in $S^2_{\TT}(T\Sigma,g)$.

We are interested in constructing a solution
$(g,\pi)$ to the constraint equations
(\ref{eq:CosmHamconstr1})--(\ref{eq:CosmHamconstr2}) with $\tr_g \pi = \tau$ for
$\tau \in \Re$. Clearly $\pi$ of the form
\begin{equation}\label{eq:pidef}
\pi^{ab} = \pi_{\TT}^{ab} + \half \tau g^{ab}
\end{equation}
will solve (\ref{eq:CosmHamconstr2}) and every solution is of this form. 
It remains to consider the Hamiltonian
constraint (\ref{eq:CosmHamconstr1}) which can be written
$$
\frac{\tau^2}{2} - |\pi_{\TT}|_g^2 + R[g] = \CosmC .
$$
From the conformal transformation property of $\pi_{\TT}$ we get
\begin{equation}\label{eq:piTTdef}
|\pi_{\TT}|_g^2 = e^{-4\lambda}|p_{\TT}|_h^2
\end{equation}
and using equations (\ref{eq:confg}) and (\ref{eq:confR}), we see that
(\ref{eq:CosmHamconstr1}) takes the form
\begin{equation}\label{eq:lambda}
\Delta_h \lambda = \half R[h] + e^{2\lambda}\frac{\tau^2- 2\CosmC}{4}
- e^{-2\lambda} \frac{|p_{\TT}|_h^2}{2} .
\end{equation}
Existence and uniqueness of solutions to (\ref{eq:lambda}) has been proved,
see \cite{moncrief:2+1:reduction}.
We now define the map $\YY_{\tau,\sigma}$.
Let $(q,p) \in T^*\Teich(\Sigma)$ be given.
By Proposition \ref{prop:teichbasic}, Point \ref{6g-6} we may introduce
global coordinates
$$
q^{\alpha}, \qquad \alpha=1,\dots,6\genus(\Sigma)-6
$$
on $\Teich(\Sigma)$.
It is natural to view
$T^* \Teich(\Sigma)$ as the pullback under $\sigma$ of the
$\TT$ part of the cotangent bundle of $\MM_{-1}$.
The space $\MM_{-1}$ is a submanifold of $S^2(T^*\Sigma)$ and
therefore $T^*_{\TT} \MM_{-1}$ in a natural way consists of contravariant
symmetric $\TT$ tensor densities. We write a general element as
$$
p'_{\TT} = \sqrt{h} p_{\TT} ,
$$
with $p_{\TT} \in S^2_{\TT}(T\Sigma,h)$.
Let $h(q^{\alpha}) = \sigma(q^{\alpha})$, then
$$
\frac{\partial h_{ab}}{\partial q^{\alpha} } =
\sigma_* \frac{\partial}{\partial q^{\alpha}} .
$$
We may now introduce coordinates in the fiber of $T^*\Teich(\Sigma)$ by
\begin{equation}\label{palphadef}
p^{\alpha} = \int_{\Sigma} p_{\TT}^{ab} \frac{\partial h_{ab}}{\partial
q^{\alpha} } \sqrt{h} d^2 x ,
\end{equation}
see \cite{moncrief:2+1:reduction} .

Conversely, given $(q,p) \in T^*\Teich(\sigma)$, with $h = \sigma(q)$,
we can define $p_{\TT}$ by the conditions
$$
p^{\alpha} =
\int_{\Sigma} p^{ab}_{\TT} \frac{\partial h_{ab}}{\partial q^{\alpha} }
\sqrt{h} d^2 x , \qquad \alpha = 1,\dots, 6 \genus(\Sigma) -6.
$$
Solve for $\lambda$ using (\ref{eq:lambda}) and	 define
$\pi_{\TT}$ by (\ref{eq:piTTdef}).
Finally, $(g,\pi)$ are defined by $g = e^{2\lambda} h$ and $\pi$ is given by
(\ref{eq:pidef}).
Now we put
\begin{equation}\label{eq:Ydef}
\YY_{\tau,\sigma}(q,p) = (g,\pi) .
\end{equation}

\subsection{Estimates for the evolution problem}
\label{sec:est-evol}
Let $(g(\tau), \pi(\tau))$ be a solution to the evolution problem in CMC gauge,
(\ref{eq:dtgpi})--(\ref{eq:dtpi}) starting at $\tau_0$, with lapse satisfying
(\ref{eq:CMCcond}) and with shift vectorfield $X$ chosen so that the
conformal metric $h(\tau) \in \sigma$ for all $\tau$. Here $h(\tau)$ is
the unique element in $\MM_{-1}$ in the $\PP$-orbit of $g(\tau)$. See
\cite{moncrief:2+1:reduction} for a discussion of the choice of $X$. The
shift vector field is chosen as the solution of an elliptic system with
coefficients in $h$ and is therefore estimated in terms of $h$.

If $\CosmC > 0$ we assume $\tau_0 > \sqrt{2\CosmC}$.
By standard theory we have short time existence for this system, in
particular there exist $\taumin < \tau_0 < \taumax $
so that we have solution for
$\taumin < \tau < \taumax$.

\begin{lem}[Area estimate]\label{lem:area}
Let $\tau_0,(g(\tau),\pi(\tau))$, $\taumin$, $\taumax$
be as above.
The area of $\Sigma$ w.r.t. $g$ is given by
\begin{equation}\label{eq:Area}
\Area(\Sigma,g) = \int_{\Sigma} \sqrt{g} d^2 x .
\end{equation}
Then for $\tau \in (\taumin,\taumax)$
\begin{eqnarray}
\label{eq:detg-estimate1}
\frac{\tau_0^2 - 2\CosmC}{\tau^2 - 2\CosmC} &\leq&
\frac{\Area(\Sigma,g(\tau))}{\Area(\Sigma,g(\tau_0))} \leq
\left ( \frac{\tau_0^2 - 2\CosmC}{\tau^2 - 2\CosmC} \right )^{1/2} ,
\qquad \text{ if $\tau > \tau_0$} , \\
\label{eq:detg-estimate2}
\left ( \frac{\tau_0^2 - 2\CosmC}{\tau^2 - 2\CosmC} \right )^{1/2} &\leq&
\frac{\Area(\Sigma,g(\tau))}{\Area(\Sigma,g(\tau_0))} \leq
 \frac{\tau_0^2 - 2\CosmC}{\tau^2 - 2\CosmC} ,
\qquad \text{ if $\tau < \tau_0$} .
\end{eqnarray}
\end{lem}
\begin{pf}
We compute $\partial_\tau \Area(\Sigma,g)$.
By assumption we are using CMC gauge, so $\tr_g \pi = \tau$. We compute
\begin{eqnarray*}
\partial_\tau \sqrt{g} &=&
\half \tr_{g} (\partial_\tau g ) \sqrt{g} \\
&=& \half \tr_{g}( -2 \Lapse (g \tr_g \pi   - \pi)  + \Lie_X g) \sqrt{g} \\
&=& (- \Lapse \tau + \div_{g} X ) \sqrt{g} .
\end{eqnarray*}
Since $\Sigma$ is compact,
$$
\int_{\Sigma}\div_{g} X	 \sqrt{g}d^2 x = 0
$$
and hence
$$
\partial_\tau \Area(\Sigma,g) = - \int_{\Sigma} \Lapse
\tau \sqrt{g} d^2 x ,
$$
which using the inequality (\ref{eq:Lapseest}) for the lapse $\Lapse$
gives the differential inequality for the area
$$
- \frac{2\tau}{\tau^2 - 2\CosmC} \leq
\partial_\tau \log( \Area(\Sigma,g) )
\leq - \frac{\tau}{\tau^2 - 2 \CosmC}.
$$
Solving this inequality gives the result.
\end{pf}

\begin{lem}[Energy estimate]\label{lem:Dirichlet}
Let $\tau_0,(g(\tau),\pi(\tau)),\taumin,\taumax$
be as above.
\begin{eqnarray}
\label{eq:hE-estimate1}
\hE(g(\tau)) &\leq& \hE(g(\tau_0))
\left (
\frac{\tau_0 + \sqrt{ \tau_0^2 - 2 \CosmC}}{\tau + \sqrt{\tau^2 - 2\CosmC}}
\right )^{\sqrt{2}}, \qquad \text{ if $\tau < \tau_0$} , \\
\label{eq:hE-estimate2}
\hE(g(\tau)) &\leq& \hE(g(\tau_0))
\left (
\frac{\tau + \sqrt{ \tau^2 - 2 \CosmC}}{\tau_0 + \sqrt{\tau_0^2 - 2\CosmC}}
\right )^{\sqrt{2}}, \qquad \text{ if $\tau > \tau_0$}	.
\end{eqnarray}
\end{lem}
\begin{pf}
By the diffeomorphism invariance of $\hE$, $D \hE(g)\Lie_X g =0$ and thus we
have by (\ref{eq:DhE}), noting that the trace free part of 
$-2N(\tr_g \pi - \pi)$ is just $2N \pi_{\TT}$,
$$
\partial_{\tau} \hE(g) =
  \int_\Sigma  \Lapse \la  \pi_{\TT}^{\sharp} \nabla S^{\ell} ,
\nabla S^{\ell}	 \ra_g \sqrt{g} d^2 x ,
$$
which gives the estimate
$$
| \partial_{\tau} \hE(g)|  \leq \max(\Lapse |\pi_{\TT}|_g) \hE(g) .
$$
Lemma \ref{lem:Lapsepiest} gives pointwise bounds on
$\Lapse$ and $\pi_{\TT}$ and we get the following differential
inequality for $\hE$,
$$
| \partial_\tau \hE(g(\tau)) | \leq
\frac{\sqrt{2}}{\sqrt{\tau^2 -2\CosmC}} \hE(g(\tau)) .
$$
Integrating this differential inequality completes the proof of the Lemma.
\end{pf}

\subsection{Proof of Theorem \ref{thm:global}}
\label{sec:proof-global}
The reduction of 2+1 gravity in the CMC gauge using the slice $\sigma$
yields a smooth time dependent Hamiltonian system on $T^* \Teich(\Sigma)$.
First we prove this has global existence in
time. Then we reconstruct the solution curve in $T^* \MM$, thus
proving global existence for the original system.
Since we are considering a smooth finite dimensional time dependent 
Hamiltonian system, we need only prove that the data do not blow up
for $\tau$ satisfying the conditions of Point \ref{point:glob1}.

Let $\tau_0$ satisfying the conditions of Point \ref{point:glob1} be given
and let $(q(\tau),p(\tau))$ be a solution
curve with $(q(\tau_0),p(\tau_0)) = (q_0,p_0)$ and let $(g(\tau),\pi(\tau)) =
\YY_{\tau}(q(\tau), p(\tau))$ be the corresponding solution curve in
$T^*\MM$. From standard theory it follows that we have existence for some
interval $(\taumin, \taumax)$ containing $\tau_0$.

From Lemma \ref{lem:Dirichlet} we know that
the Dirichlet energy $\tE(q)$ is bounded for $\tau \in (\taumin,\taumax)$
By Proposition \ref{prop:tromba}, the Dirichlet energy is a proper function on
$\Teich(\Sigma)$ and therefore for such $\tau$,
$q(\tau)$ stays in a compact subset of $\Teich(\Sigma)$.

Now consider $h(\tau) = \sigma(q(\tau))$. By the construction of the slice
$\sigma$ we have under the above conditions on $\tau$ uniform
pointwise estimates for $h(\tau)$. In the following we suppress reference to
$\tau$.

Let $g = e^{2\lambda} h$ with $h \in \MM_{-1}$. Then by the considerations in
subsection \ref{sec:confconstr} we have
$\pi^{ab}_{\TT} = e^{-4\lambda} p^{ab}_{\TT} $.
We estimate the coordinates in the fiber of $T^* \Teich(\Sigma)$ using
(\ref{palphadef}) and the H\"older inequality,
\begin{eqnarray*}
| p^{\alpha} | &=&
\left \vert \int_{\Sigma} p^{ab}_{\TT} \frac{\partial h_{ab}}{\partial q^{\alpha} }
\sqrt{h} d^2 x \right \vert \\
&=& \left \vert	 \int e^{-\lambda} p^{ab}_{TT}
\frac{\partial h_{ab}}{\partial q^{\alpha}}e^{\lambda}	\sqrt{h} d^2 x \right
\vert \\
&\leq& \left \Vert \frac{\partial h}{\partial q^{\alpha}} \right \Vert_{h,\max}
\left ( \int_\Sigma e^{2\lambda} \sqrt{h} d^2 x \right )^{1/2}
\left ( \int_{\Sigma} e^{-2\lambda} p_{\TT}^{ab} p_{\TT}^{cd} h_{ac} h_{bd} \sqrt{h} d^2 x
\right )^{1/2} \!\!\!.
\end{eqnarray*}
where
$\left \Vert \frac{\partial h}{\partial q^{\alpha}} \right \Vert_{h,\max}$
is the maximum modulus w.r.t. $h$ of the tensor
$\frac{\partial h}{\partial q^{\alpha}}$. By construction this is uniformly
bounded for $\tau$ satisfying our assumptions. Further
$e^{2\lambda} \sqrt{h} =\sqrt{g}$ so
$$
\int_\Sigma e^{2\lambda} \sqrt{h} d^2 x	 = \Area(\Sigma,g) ,
$$
which is estimated in Lemma \ref{lem:area}. It remains to consider the last
factor. Note that by the conformal transformation rules,
$$
e^{-2\lambda}  p_{\TT}^{ab} p_{\TT}^{cd} h_{ac} h_{bd} \sqrt{h} =
\pi_{\TT}^{ab} \pi_{\TT}^{cd}
g_{ac} g_{bd} \sqrt{g} = | \pi_{\TT}|^2_g \sqrt{g}.
$$
Using (\ref{eq:tKest}) we  now have the bound
$$
|p^{\alpha}| \leq \left \Vert \frac{\partial h}{\partial q^{\alpha}} \right
\Vert_{h,\max} \Area(\Sigma, g)
\left ( \frac{\tau^2 - 2 \CosmC}{2} \right)^{1/2} \!.
$$
Referring again to the area estimate in Lemma \ref{lem:area} we find that the
coordinates $p^{\alpha}$ in the fiber of $T^* \Teich(\Sigma)$ are bounded
uniformly under the conditions on $\tau$.
Therefore it follows that we can extend the interval of existence for the
reduced system to the intervals claimed in Point \ref{point:glob1}.

We have now proved global existence for the reduced version of 2+1 vacuum
GR. It remains to reconstruct the solution curve. This is done using the 
map $\YY_{\tau,\sigma}$. By construction this is a smooth map. The lapse and shift are
governed by elliptic equations which satisfy uniform estimates for $\tau$
satisfying the present assumptions. This finishes the proof of Point
\ref{point:glob1}. Point \ref{point:glob2} follows from Lemma \ref{lem:area}.

\end{document}